\documentclass[12pt]{elsarticle}
\usepackage{graphicx}
\usepackage{amssymb}
\usepackage{amsmath}
\usepackage{calc}

\usepackage{fullpage}

\newcommand{\proof}{{\noindent}{\bf Proof.\ \ }}
\newcommand{\gp}{\psi}
\newcommand{\rp}{\psi_r}

\newtheorem{theorem}{Theorem}
\newtheorem{lemma}[theorem]{Lemma}

\newtheorem{cor}[theorem]{Corollary}
\newtheorem{obs}[theorem]{Observation}

\begin{document}

\begin{frontmatter}

\title{On polygon numbers of circle graphs and distance hereditary graphs}

\author[ls]{Lorna Stewart\corref{correspondingauthor}}
\cortext[correspondingauthor]{Corresponding author}
\ead{lorna.stewart@ualberta.ca} 

\author[rv]{Richard Valenzano}
\ead{rvalenzano@cs.toronto.edu}

\address[ls]{Department of Computing Science, University of Alberta, Edmonton, Alberta, Canada T6G 2E8}
\address[rv]{Department of Computer Science, University of Toronto, Toronto, Ontario, Canada M5S 3G4}

\date{Sept 7, 2017}

\begin{abstract}
Circle graphs are intersection graphs of chords in a circle and $k$-polygon graphs are intersection graphs of chords
in a convex $k$-sided polygon where each chord has its endpoints on distinct sides.
The $k$-polygon graphs, for $k \ge 2$, form an infinite chain of graph classes, each of which contains the class of permutation graphs.
The union of all of those graph classes is the class of circle graphs. 
The polygon number $\gp(G)$ of a circle graph $G$ is the minimum $k$ 
such that $G$ is a $k$-polygon graph.
Given a circle graph $G$ and an integer $k$, determining whether $\gp(G) \le k$ is NP-complete,
while the problem is solvable in polynomial time for fixed $k$.

In this paper, 
we show that $\gp(G)$ is always at least as large as the asteroidal number of $G$, and equal to the asteroidal number of $G$ when $G$ is a connected distance hereditary graph that is not a clique.
This implies that the classes of distance hereditary permutation graphs and distance hereditary AT-free graphs are the same, and we give a forbidden subgraph characterization of that class.
We also establish the following upper bounds:
$\gp(G)$ is at most the clique cover number of $G$ if $G$ is not a clique,
at most 1 plus the independence number of $G$, 
and at most $\lceil n/2 \rceil$ where $n \ge 3$ is the number of vertices of $G$.
Our results lead to linear time algorithms for finding the 
minimum number of corners that must be added to a given circle representation to produce a polygon representation,
and for finding the asteroidal number of a distance hereditary graph, both of which are improvements over previous algorithms for those problems.
\end{abstract}


\end{frontmatter}

\section{Introduction and Preliminaries}\label{intro}

The dimension of comparability graphs and the treewidth of graphs are widely studied graph parameters that are important from both algorithmic and structural points of view \cite{Trotter,treewidth}.
In this paper, we study an analogous parameter of circle graphs, namely, the polygon number.
The three parameters have similar algorithmic and complexity properties, and each of them may be seen as a parameter of an associated representation: a realizer of a partially ordered set, a tree decomposition of a graph, or a polygon representation of a circle graph.
Further similarities between the polygon number of a circle graph and the dimension of a comparability graph will be mentioned later.

The $k$-polygon graphs, for $k \ge 2$, form an infinite chain of graph classes, each of which contains the class of permutation graphs, and
the union of which is the class of circle graphs. 
The polygon number of a given circle graph is the minimum value of $k$ such that the graph is a $k$-polygon graph.
Given a circle graph $G$ and an integer $k$, determining whether the polygon number of $G$ is at most $k$ is NP-complete, while the problem is solvable in polynomial time for fixed $k$ \cite{ES}.
Several problems that are known to be NP-hard on circle graphs admit polynomial time algorithms for $k$-polygon graphs when $k$ is fixed, including 
domination and independence problems, the topological via minimization problem in circuit design \cite{ES2}, and vertex colouring with a fixed number of colours \cite{Unger92}. In addition, small collective additive tree spanners can be constructed efficiently for $k$-polygon graphs when $k$ is fixed \cite{DCKX},
and the bandwidth of a $k$-polygon graph can be approximated to within a factor of $2k^2$ in polynomial time \cite{KratschStewart}.
The running times of several of these algorithms are of the form $O(f(|V|) \cdot |V|^{g(k)})$ where $V$ is the vertex set of the input graph and $f$ and $g$ are polynomial functions. 

Although the polygon number has been a key parameter in algorithm design, and the complexity of computing it is known, little is known about its other properties.
In this paper we explore 
how the polygon number of a circle graph relates to established graph parameters. 
This gives some insight into how the $k$-polygon graph classes increase in complexity as $k$ increases, and provides estimates on the running times and approximation ratios of algorithms for $k$-polygon graphs.
Specifically, we
show that the polygon number is at least as large as the asteroidal number, with equality for connected distance hereditary graphs other than cliques.
This implies that the classes of distance hereditary permutation graphs and distance hereditary AT-free graphs are the same, and leads to a forbidden subgraph characterization of that class.
We then show that the polygon number of a circle graph is
at most the clique cover number (if the graph is not a clique),
at most 1 plus the independence number,
and at most 
$\lceil n/2 \rceil$ where $n \ge 3$ is the number of vertices of the graph.
These results give rise to 
linear time algorithms for computing the minimum number of corners that must be added to a given circle representation to construct a polygon representation, and for computing the asteroidal number of a distance hereditary graph.

We begin with terminology and preliminaries, first for graphs and then for intersection representations of circle and $k$-polygon graphs. Additional definitions and notation are introduced as needed. For terms not defined here, the reader is referred to \cite{Golumbic}. 

The graphs that we consider are finite and simple, and undirected unless stated otherwise.
When the vertex and edge sets of a graph $G$ are not explicitly named, we refer to them as $V(G)$ and $E(G)$, respectively.
Let $G=(V,E)$ be a graph. 
The subgraph of $G$ induced by $W \subseteq V$ is denoted $G[W]$.
For $v \in V$, the {\em neighbourhood} of $v$ is $N_G(v) = \{ w ~|~ vw \in E \}$, the {\em closed neighbourhood} of $v$ is $N_G[v] = N(v) \cup \{v\}$, and the {\em degree} of $v$ is denoted $d_G(v)$. The neighbourhood of a subset $W$ of $V$ is $N_G(W)= \{ v \in V - W ~|~ wv \in E \mbox{ for some } w \in W \}$. 
The subscript $G$ may be omitted when the context is clear.
We use $G - V'$ and $G - E'$ as shorthand for the subgraph of $G$ induced by $V - V'$, and the graph $(V, E - E')$, respectively. 
A vertex of degree one is called a {\em leaf}. 
The chordless cycle on $n$ vertices and the clique on $n$ vertices are denoted by $C_n$ and $K_n$ respectively.
The size of a maximum independent set is denoted $\alpha(G)$ and the size of a minimum clique cover is denoted $\kappa(G)$.
Using the notation of \cite{aster}, a set $A \subseteq V$ is called an {\em asteroidal set} if for every vertex $a \in A$, there is a path between each pair of vertices $x, y \in A - \{a\}$ in $G - N[a]$. 
The {\em asteroidal number}, denoted $an(G)$, of $G$ is the cardinality of a maximum asteroidal set of $G$. An {\em asteroidal triple} (or {\em AT}) is an asteroidal set of size three.

A graph is called {\em AT-free} if it has no asteroidal triple. A graph is a {\em comparability graph} if its edges can be transitively oriented, and a {\em cocomparability graph} if it is the complement of a comparability graph. A graph $G$ is a {\em distance hereditary graph} if,
for every connected induced subgraph $H$ of $G$, the distance between each pair of vertices of $H$ is the same in $H$ as it is in $G$.
We refer the reader to \cite{GraphClasses} for more information about these graph classes.

The {\em intersection graph} of a finite collection of sets is the graph containing one vertex for each set, such that two vertices are joined by an edge if and only if the intersection of the corresponding sets is not empty. 

A graph is a {\em circle graph} if it is the intersection graph of a set of chords of a circle.
For $k \geq 3$, a graph is a {\em $k$-polygon graph} if it is the intersection graph of chords inside a convex polygon with $k$ sides such that each chord has its endpoints on two distinct sides of the polygon.
For example, for any $k \ge 3$, $C_{2k}$ is a $k$-polygon graph and not a ($k$-1)-polygon graph \cite{ES}.
A graph $G$ where $V(G) = \{ v_1, \ldots, v_n \}$ is a permutation graph if there exists a permutation $\pi$ of $\{ 1,2, \ldots, n\}$ such that $v_i v_j \in E(G)$ if and only if $(i-j) \cdot (\pi^{-1}_i - \pi^{-1}_j ) <0$. Equivalently, permutation graphs are the intersection graphs of straight line segments connecting two parallel lines.
For reasons that will be made evident below, permutation graphs are considered to be $2$-polygon graphs.
Therefore:
\[ \mbox{permutation graphs} \equiv \mbox{2-polygon graphs} \subset \mbox{3-polygon graphs} 
\subset \ldots \]
\[\ldots \subset \sum_{k=2}^\infty \mbox{$k$-polygon graphs} \equiv \mbox{circle graphs} \]

The {\em polygon number} of a circle graph $G$, $\gp(G)$ is the minimum value of $k$ such that $G$ is a $k$-polygon graph.
In \cite{ES}, Elmallah and Stewart showed that the problem of determining if $\gp(G) \le k$ for a given circle graph $G$ and an integer $k$ is NP-complete, and
they gave a polynomial time algorithm for solving the problem when $k$ is a fixed integer. They also showed that
for a circle graph $G$ with connected components
$G_1, G_2, \ldots, G_r$, 
\[ \gp(G) = \left( \sum_{i=1}^r \gp(G_i) \right) - 2(r-1). \]
As this allows us to determine $\gp(G)$ based on the polygon numbers of the connected components of $G$, we focus on identifying the polygon number of connected graphs in the analysis below.

\begin{figure}
\centering
\includegraphics[scale=.95]{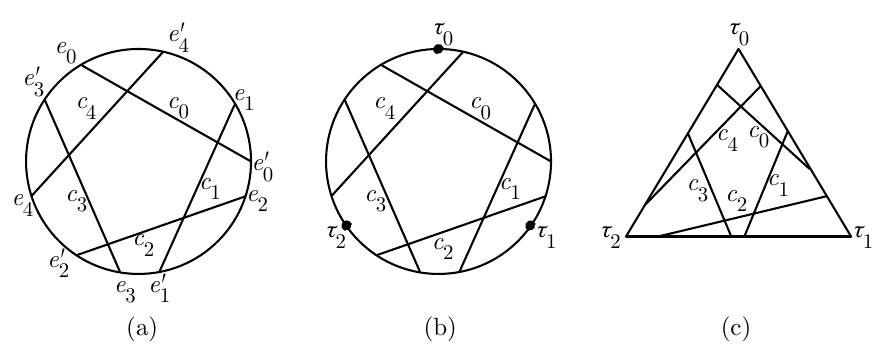}
\caption{Examples of circle and polygon representations}\label{RepExample}
\end{figure}

A set of chords of a circle is called a {\em circle representation} for graph $G$ if $G$ is the intersection graph of that set of chords. For example, Figure \ref{RepExample}(a) shows a circle representation for $C_5$ where each chord $c_i$ has endpoints at points labelled as $e_i$ and $e_i'$. 
Two distinct points $p_0$ and $p_1$ divide the circle into the two arcs: $(p_0,p_1)$, the open arc that is traced in a clockwise traversal of the circle beginning at $p_0$ and ending at $p_1$, and $(p_1,p_0)$ which is defined analogously.
For chord $c$ with endpoints at points $e$ and $e'$ we say that $(e,e')$ and $(e',e)$ are the {\em arcs of $c$}.
An arc of the circle is defined to be {\em empty} if it does not contain both endpoints of any chord.
A chord that has an empty arc is said to be {\em peripheral} and we use ${\mathcal C}_{\emptyset}$ to denote the set of peripheral chords of a set of chords $\mathcal C$.
For example, all chords shown in Figure \ref{RepExample}(a) are peripheral, and of the arcs of $c_0$, $(e_0, e_0')$ is empty and $(e_0', e_0)$ is not empty.
These definitions imply the next two observations.

\begin{obs}\label{uv}
A chord has two empty arcs in a circle representation of graph $G$ if and only if it represents a universal vertex in $G$. 
\end{obs}

\begin{obs}
\label{point_in_common}
The empty arcs of two distinct peripheral chords in a circle representation contain a common point if and only if the chords intersect.
\end{obs}

A {\em $k$-polygon representation} of a graph $G$ is a set of chords of a $k$-sided convex polygon where each chord has its endpoints on two distinct sides, such that $G$ is the intersection graph of the chords.
A circle representation for graph $G$ can be transformed into a $k$-polygon representation for $G$ (for large enough $k$) by
adding a set of $k$ points representing the corners of the polygon. These points will be referred to as {\em corners}.
A chord is {\em satisfied} by the added corners if it has a corner in each of its arcs. 
If all chords in the representation are satisfied then we call the added corners a {\em satisfying set of corners} and the polygon representation can then be constructed by straightening the arcs between consecutive corners. 
This means that for any two of the added corners, $\tau_0$ and $\tau_1$, such that $(\tau_0, \tau_1)$ does not contain any other corner, the arc will be replaced by a straight line on which the relative ordering of endpoints remains the same. 
In doing so, the set of chord crossings in the representation does not change.
Applying this transformation for each such pair of neighbouring corners results in a $k$-polygon representation for $G$.
For this reason, we will often view a $k$-polygon representation as a circle representation with the addition of $k$ corners.
For example, Figure \ref{RepExample}(b) shows a $3$-polygon representation of $C_5$ which was constructed by adding three corners -- labelled $\tau_0$, $\tau_1$, and $\tau_2$ -- to the circle representation shown in Figure \ref{RepExample}(a).
$ \lbrace \tau_0, \tau_1, \tau_2 \rbrace$ is a satisfying set of corners.
Figure \ref{RepExample}(c) shows the result of the arc straightening procedure described above.

A {\em permutation representation} for a graph $G$ is a set of straight line segments connecting two parallel lines, the intersection graph of which is $G$.
If a satisfying set of corners for a circle representation contains only two distinct corners $\tau_0$ and $\tau_1$, then a permutation representation can be constructed by replacing $(\tau_0, \tau_1)$ and $(\tau_1, \tau_0)$ by parallel lines and maintaining the relative ordering of the endpoints on these new lines.
It is this transformation and its inverse that allow us to refer to permutation graphs as $2$-polygon graphs. 

Without loss of generality, we assume that chord and line segment endpoints are distinct in circle, $k$-polygon, and permutation representations, and that chord endpoints are distinct from the corners in $k$-polygon representations. 

The next two observations follow from the above definitions.

\begin{obs}
\label{sub}
For any circle graph $G$ and any induced subgraph $H$ of $G$, 
$\gp(H) \le \gp(G)$.
\end{obs}

\begin{obs}
\label{sat_periph}
A set of corners satisfies the peripheral chords of a representation if and only if it satisfies all chords of the representation.
\end{obs}

Different circle representations for a graph may require different numbers of corners to be added to produce polygon representations.
The {\em polygon number of a circle representation} $\mathcal{C}$, denoted by $\rp(\mathcal{C})$, is the minimum number of corners that must be added to $\mathcal{C}$ in order to satisfy all chords of $\mathcal{C}$.

\begin{obs}\label{minoverallreps}
For every circle graph $G$, 
\[
\gp(G) ~= ~
\min_{\mathcal{C}}
\ \rp(\mathcal{C})
\]
where the minimum is taken over all circle representations $\mathcal{C}$ of $G$.
\end{obs}

A quadratic algorithm to compute the polygon number of a circle representation is given in \cite{ES}.
In Section \ref{algorithms}, we describe a linear time algorithm for this problem.

We now introduce concepts regarding the relative positions of chord endpoints and corners in circle and polygon representations.
Given $\ell$ distinct points on a circle, 
we use 
$\langle p_0, p_1 , \ldots , p_{\ell - 1} \rangle$ to mean that in a clockwise traversal of the circle beginning at point $p_0$ and ending before $p_0$ is encountered a second time, points $p_0, \ldots, p_{\ell - 1}$ are encountered in that order.
Note that points other than $p_0, \ldots, p_{\ell - 1}$ may also be encountered during the traversal.
For example, in Figure \ref{RepExample}(a), $e_0'$, $e_1$, $e_3$, and $e_4'$ may be referred to as being in the order given by $\langle e_4', e_1 , e_0', e_3 \rangle$.
We say that a chord is {\em in a corner} $\tau$ if its endpoints are on the two sides that meet at $\tau$.
If $k=2$ then all chords are in both corners. 
For $k>2$, each chord is in at most one corner.
A chord endpoint $e$ is said to be {\em close to a corner} $\tau$ if $e$ is on one of the sides that meet at $\tau$ and there is no other endpoint on that side between $e$ and $\tau$.
We say that a chord $c$ is {\em close to a corner} $\tau$ if at least one of its endpoints is close to  $\tau$. An {\em extreme chord} in a polygon representation is one that is close to a corner.
For example, in Figure \ref{RepExample}(c), chord $c_1$ is in $\tau_1$ and is close to $\tau_1$; $c_2$ is close to both $\tau_1$ and $\tau_2$. All chords shown in Figure \ref{RepExample}(c), except $c_3$, are extreme.

Finally, a sequence $c_0,c_1, \ldots, c_{\ell-1}$ of $\ell \ge 1$ chords in a circle is an {\em $\ell$-independent set in series} if there is a point $x$ on the circle, such that,
in a clockwise traversal of the circle starting at $x$ and ending before $x$ is encountered a second time, both endpoints of $c_i$ are encountered before both endpoints of $c_{i+1}$ for all $0 \le i < \ell-1$.

\section{Asteroidal number and distance hereditary graphs}\label{DH}

In this section, we show that the asteroidal number of a circle graph is a lower bound on the polygon number, and that equality holds for every connected distance hereditary graph that is not a clique. We also identify a forbidden subgraph characterization of distance hereditary permutation graphs.

\begin{theorem}
\label{k_bound_by_an}
For any circle graph $G$, $\gp(G) \ge an(G)$.
\end{theorem}

\proof
If $an(G) \le 2$, the statement is true since $\gp(G) \geq 2$ for all circle graphs. 
Now consider the case where $an(G) \geq 3$.
Let $\mathcal{C}$ be a $\gp(G)$-polygon representation of $G$, $A$ be an asteroidal set of $G$ where $|A| = an(G)$, and $\mathcal{C}_A \subseteq \mathcal{C}$ be the set of chords corresponding to vertices in $A$.

We first show that $\mathcal{C}_A$ forms an independent set in series.
Consider any vertex $v \in A$, corresponding to chord $c_v$,
and let $a, b \in A - \{v\}$ be two other vertices of $A$ that correspond to chords $c_a$ and $c_b$, respectively. 
By the definition of $A$, neither intersects $c_v$. 
Therefore, both endpoints of $c_a$ must be in the same arc of $c_v$. 
The same is true for the endpoints of $c_b$.
Suppose that the endpoints of $c_a$ are in one arc while the endpoints of $c_b$ are in the other arc of $c_v$.
Then at least one vertex of any path from $a$ to $b$ must correspond to a chord with one endpoint in each arc of $c_v$.
Such a chord intersects $c_v$, and so the corresponding vertex is in $N(v)$, which contradicts that $A$ is an asteroidal set.
Therefore, the endpoints of $c_a$ and the endpoints of $c_b$ are contained in the same arc of $c_v$. This shows that for every $c_v \in \mathcal{C}_A$, the endpoints of all other chords of $\mathcal{C}_A$ are in the same arc of $c_v$. Let $c_v$ be a chord of $ \mathcal{C}_A$ with endpoints $e_v$ and $e'_v$ such that the arc $(e_v , e'_v)$ contains no endpoints of chords of $\mathcal{C}_A$. 
If $(e_v , e'_v)$ is not empty, we can obtain another asteroidal set of size $|A|$ by replacing $c_v$ with a peripheral chord that has both endpoints in $(e_v , e'_v)$. Thus we assume without loss of generality that the chords of $\mathcal{C}_A$ are all peripheral.
Now, the order in which the endpoints of chords of $ \mathcal{C}_A$ are encountered in a clockwise traversal of the circle beginning at $e_v$ satisfies the definition of an $|A|$-independent set in series, that is, 
$\mathcal{C}_A$ forms an $|A|$-independent set in series.

Therefore, since each chord of $\mathcal{C}_A$ must have a corner in its empty arc, and by Observation \ref{point_in_common}, $\gp(G) = \rp(\mathcal{C}) \geq |A| = an(G)$.
$\Box$
\bigskip

\begin{figure}
\centering
\includegraphics{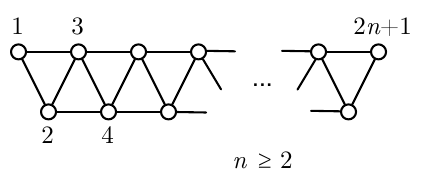}
\caption{An infinite family of unit interval graphs with asteroidal number two and arbitrarily large polygon number}\label{uig-an}
\end{figure}

In general the gap between $\gp(G)$ and $an(G)$ is unbounded. For example, the unit interval graphs depicted in Figure \ref{uig-an} have asteroidal number two and arbitrarily large polygon number.

However, in Theorem \ref{dh} we show that $\gp(G) = an(G)$ when $G$ is a connected distance hereditary graph that is not a clique. The development of that theorem relies on split decompositions, the subject of the next several definitions and lemmas.

A {\em module} in a graph $G$ is a subset $M$ of $V(G)$ such that each vertex of $V(G) - M$ is adjacent to all or to none of the vertices of $M$. All of the modules referred to in this paper are nontrivial, that is, $1 < |M| < |V(G)|$.
Two vertices are called {\em twins} if they form a module of size two.
If $M$ is a module of $G$ and $v$ is a vertex of $M$, then the operation of {\em reducing module $M$ to $v$} in $G$ results in the graph $G - ( M - \{v\} )$.
Given graphs $G$ and $H$ with disjoint nontrivial vertex sets, and a vertex $v \in V(G)$, the operation of {\em substituting $v$ with $H$} in $G$ produces the graph obtained by taking the union of
$G- \{v\}$ and $H$ and adding all edges between 
$N_G(v)$ and $V(H)$. Note that $V(H)$ is a module in the resulting graph.
If $M$ is a module of $G$ such that $G[M]$ has a certain property, we also say that $M$ has the property. 

A {\em split} of a connected graph $G$ is a partition $\{V_1, V_2\}$ of $V(G)$ such that 
$|V_1| >1$, $|V_2| >1$, and $E(G) \cap \{ xy~ |~ x \in V_1 \mbox{ and } y \in V_2 \} =  \{ xy~ | ~x \in N(V_2) \mbox{ and } y \in N(V_1) \}$.
A graph with no split is called {\em prime}. Every partition $\{V_1, V_2\}$ of the vertices of a clique or a star satisfying $|V_1| >1$ and $|V_2| >1$ is a split.

Cunningham and Edmonds \cite{Cunningham} \cite{CunEdm} defined a recursive decomposition scheme for graphs based on the concept of a split. 
If $\{V_1, V_2\}$ is a split of $G$ then 
$\{G_1, G_2 \}$ is called a {\em simple split decomposition} of $G$,
where
$G_1$ is the graph $G[V_1]$ to which a new vertex $v \notin V(G)$ has been added with $N_{G_1}(v) = N_G(V_2)$ and
$G_2$ is the graph $G[V_2]$ to which vertex $v$ has been added with $N_{G_2}(v) = N_G(V_1)$.
In the resulting graphs $G_1$ and $G_2$, $v$ is called the {\em marker vertex} of the decomposition.
A {\em split decomposition} of $G$ is $\{ G \}$ or any set of graphs obtained from a split decomposition of $G$ by replacing one graph of the decomposition with the two graphs of one of its simple split decompositions.
Each graph in a split decomposition of a connected graph is connected.
The reverse of a split decomposition is the {\em join composition}.
Given graphs $G_1$ and $G_2$, each with distinguished marker vertex $v$,
the {\em join} of $G_1$ and $G_2$ is the graph is obtained by taking the union of graphs
$G_1 - \{v\}$ and $G_2 - \{v\}$ and adding the edges
$\{ xy~|~x \in N_{G_1}(v) \mbox{ and } y \in N_{G_2}(v) \}$.
Applying the join composition repeatedly in any order to a split decomposition for $G$ reconstructs the graph $G$.

Every connected graph has a unique split decomposition, called the {\em standard split decomposition}, that contains only prime graphs, cliques, and stars and is not a strict refinement of any other such split decomposition \cite{Cunningham}.
This means that applying the join composition to two graphs of a standard split decomposition that share a marker vertex does not result in a clique or a star. Equivalently, no two cliques share a marker, and if two stars share a marker then the marker is either a leaf of both stars or the centre of both stars.

Split decomposition plays a key role in the study of circle graphs. A graph is a circle graph if and only if every graph of its standard split decomposition is a circle graph, and prime circle graphs have unique circle representations up to rotation and reflection \cite{Bouchet} \cite{GHS}. Circle representations for two graphs can be combined in a straightforward way to construct a circle representation for the join composition of the two graphs. We next show how to combine two polygon representations in a similar way.

\begin{lemma}\label{Lcombine}
Let $G_1$ and $G_2$ be the graphs of a simple split decomposition of a circle graph $G$, each with marker vertex $v$.
If $G_1$ has a $k_1$-polygon representation and $G_2$ has a $k_2$-polygon representation then $G$ has a $(k_1 + k_2)$-polygon representation.
Furthermore, if $G_1$ has a $k_1$-polygon representation in which $v$'s chord is close to $q$ corners then $G$ has a $(k_1 + k_2 - q)$-polygon representation.
\end{lemma}

\begin{figure}
\centering
\includegraphics[width=\linewidth]
{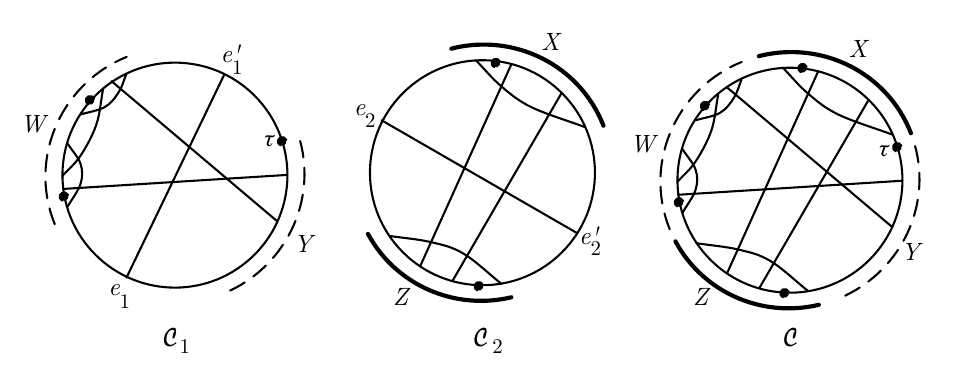}
\caption{$\mathcal{C}_1$ and $\mathcal{C}_2$ are combined to produce $\mathcal{C}$, a polygon representation for the join of the graphs represented by $\mathcal{C}_1$ and $\mathcal{C}_2$, as described in Lemma \ref{Lcombine}. In $\mathcal{C}_1$, the marker vertex corresponds to the chord with endpoints at $e_1$ and $e_1'$, while it has endpoints $e_2$ and $e_2'$ in $\mathcal{C}_2$. The black dots in the figures represent corners.}\label{combine}
\end{figure}

\proof
Let $\mathcal{C}_1$ and $\mathcal{C}_2$ be $k_1$ and $k_2$-polygon representations for $G_1$ and $G_2$, respectively. 
Let $e_1$ and $e'_1$ be the endpoints of $v$'s chord in $\mathcal{C}_1$ and let
$e_2$ and $e'_2$ be the endpoints of $v$'s chord in $\mathcal{C}_2$.
Let $W$ and $Y$ be the sequences of endpoints and corners that appear in the arcs
$(e_1, e'_1)$ and $(e'_1, e_1)$ of $\mathcal{C}_1$ respectively and let
$X$ and $Z$ be defined analogously for the arcs
$(e_2, e'_2)$ and $(e'_2, e_2)$ of $\mathcal{C}_2$.
Then the circular sequence of endpoints and corners given by $WXYZ$ 
is a $(k_1 + k_2 )$-polygon representation for $G$, as shown in Figure \ref{combine}.

Let $\mathcal{C}$ be a $(k_1 + k_2 )$-polygon representation for $G$ constructed in this manner.
We now show that all of the chords of $\mathcal{C}$ will still be satisfied after the removal of all of the corners that $v$'s chord is close to in $\mathcal{C}_1$. Notice that the chords of $\mathcal{C}_2$ are still satisfied by corners that were originally in $\mathcal{C}_2$ and any chord of $\mathcal{C}_1$ that contained only close-to-$v$ corners in one or both of its arcs now contains corners of $\mathcal{C}_2$ in those arcs since there must be at least one corner in each arc of $v$'s chord in $\mathcal{C}_2$.
Therefore the corners of $\mathcal{C}_1$ that are close to $v$'s chord in $\mathcal{C}_1$ can be deleted from $\mathcal{C}$ to produce a $(k_1 + k_2 - q )$-polygon representation for $G$. 
$\Box$

\bigskip
Notice that a chord can be close to at most four corners. Furthermore, it can only be close to at most two corners if the representation has a minimum number of corners.
This is justified by noticing that if three or four corners are close to a single chord, then all but two of them can be removed while leaving all chords of the representation satisfied.
As such, the value of $q$ in Lemma \ref{Lcombine} can be at most two if $k_1 = \gp(G_1)$.
For example, notice that $q=1$ in Figure \ref{combine} and the corner that is close to the marker chord in $\mathcal{C}_1$ is labelled $\tau$.

A split decomposition of a graph can be viewed as a tree
that contains a node for each graph of the decomposition and an edge between two nodes if the graphs share a marker vertex \cite{Cunningham}.
Formally, the {\em split decomposition tree} corresponding to a split decomposition $D$ is a tree $T$ with
$|D|$ nodes that are in one-to-one correspondence with the graphs of $D$. For each node $x$ of $V(T)$ we let $G_x$ denote the corresponding graph of $D$. Then 
$E(T) = \{ xy ~|~ x,y \in V(T) \mbox{ and } G_x \mbox{ and } G_y \mbox{ share a marker vertex} \}$.
The {\em standard split decomposition tree} of a graph $G$ is the split decomposition tree corresponding to the standard split decomposition of $G$ and will be denoted $ST_G$.

It follows from the definition of the standard split decomposition that any subtree of the standard split decomposition tree of a graph 
is the standard split decomposition tree of a subgraph of the graph.

Our strategy in the remainder of this section is to 
examine the polygon numbers and asteroidal numbers of connected graphs in terms of standard split decomposition trees.
We study general graphs and circle graphs first and then apply the results later in the context of distance hereditary graphs.
As a first step, for graph $G$, we consider the effect of removing 
certain leaves from $ST_G$. Note that the graph corresponding to a leaf contains exactly one marker vertex. We maintain this property as leaves are removed by adopting the following convention: when a leaf $x$ is removed from $ST_G$, 
if $\ell$ is the marker vertex of $G_x$ and $y$ is the neighbour of $x$ in $ST_G$, then $\ell$ 
becomes an ordinary (that is, non-marker) vertex in $G_y$.

\begin{lemma}\label{one-step}
Let $G$ be a connected graph.
If $x$ is a leaf of $ST_G$ and the marker vertex $\ell$ of $ G_x$ is universal in $G_x$,
then the following hold:
\begin{enumerate}
\item $V(G_x) - \{\ell\}$ is a module in $G$.
\item The graph obtained from $G$ by reducing the module $V(G_x) - \{\ell\}$ to a vertex and renaming it $\ell$ is connected, and $ST_G - \{x\}$ is its standard split decomposition tree.
\end{enumerate}
\end{lemma}
\proof
Since $x$ is a leaf and $\ell$ is universal in $G_x$, the split of $G$ corresponding to $\ell$ shows that $V(G_x) - \{\ell\}$ is a module in $G$.
Reducing a (nontrivial) module to a single vertex does not affect whether the graph is connected; therefore the graph obtained by reducing $V(G_x) - \{\ell\}$ to a vertex in $G$ is connected.
Let $y$ be the neighbour of $x$ in $ST_G$. Replacing $G_x$ and $G_y$ in the standard decomposition of $G$ with $G_{xy}$, the join of $G_x$ and $G_y$, and then removing all but one of the vertices of $V(G_x) - \{\ell\}$ from $G_{xy}$, produces a decomposition of the graph obtained from $G$ by reducing $V(G_x) - \{\ell\}$ to a vertex. The split decomposition tree corresponding to that decomposition is the same as $ST_G - \{x\}$ except that the vertex of $ST_G - \{x\}$ corresponding to the single remaining vertex of $V(G_x) - \{\ell\}$ is labelled $\ell$. Therefore, $ST_G - \{x\}$ is a split decomposition of the graph obtained from $G$ by reducing $V(G_x) - \{\ell\}$ to a vertex and renaming it $\ell$. Furthermore, it is a standard split decomposition tree since it is a subtree of $ST_G$.
$\Box$
\bigskip

For a connected graph $G$, we define the {\em pruned standard split decomposition tree}, denoted $PST_G$, 
based on the following process: Initially, let $T$ be $ST_G$; while $T$ has a leaf $x$ such that the marker vertex of $G_x$ is universal in $G_x$, remove $x$ from $T$. Now, $PST_G$ is the tree $T$ that remains after the process terminates.
Since $PST_G$ is a subtree of $ST_G$, it is the standard split decomposition tree of a subgraph of $G$.
We will use $H_G$ to denote that subgraph of $G$, that is, the graph satisfying $ST_{H_G} = PST_G$.
The next lemmas outline relationships among $G$, $PST_G$, and $H_G$, where $G$ is a connected graph.

\begin{lemma}\label{prunesplittree}
For any connected graph $G$, $H_G$ is a connected induced subgraph of $G$ that can be obtained from $G$ by repeatedly reducing a module to a single vertex and renaming that vertex. If $G$ is a circle graph, then the reduced modules all induce permutation graphs.
\end{lemma}

\proof
The statements follow by repeated applications of Lemma \ref{one-step} and by the fact that any circle graph that has a universal vertex is a permutation graph (see \cite{Golumbic} Ex. 12, p. 252).
$\Box$

\begin{lemma}\label{psipreserved}
For any connected graph $G$, the following hold:
\begin{enumerate}
\item
$an(H_G) \le an(G)$.
\item
If $G$ is a circle graph then $\psi(H_G) = \psi (G)$.
\end{enumerate}
\end{lemma}

\proof
Part (1) holds since $H_G$ is an induced subgraph of $G$ (Lemma \ref{prunesplittree}) and the asteroidal number of any induced subgraph of $G$ is at most $an(G)$ \cite{aster}.
To justify (2), first notice that $\psi(H_G) \le \psi(G)$ by Observation \ref{sub}.
We now show that a polygon representation for $H_G$ can be extended to a polygon representation for $G$ without adding any corners.
This holds by Lemmas \ref{Lcombine} and \ref{prunesplittree}, and the observation that substituting a vertex with a permutation graph is equivalent to performing a join composition where one of the graphs is a permutation graph with a universal vertex marker, and such a graph has a 2-polygon representation in which the marker is close to two corners.
$\Box$

\begin{lemma}\label{an>=leaves}
For any connected graph $G$, $an(G)$ is greater than or equal to the number of leaves of $PST_G$.
\end{lemma}

\proof
The inequality is obviously true if $PST_G$ is trivial or has two leaves. Now suppose that $PST_G$ has three or more leaves. 
We construct an asteroidal set of $G$ consisting of one vertex from each of the graphs corresponding to the leaves of $PST_G$, as follows.
For each leaf, choose from the corresponding graph a non-marker vertex that is not adjacent to the marker (such a vertex exists by the definition of  $PST_G$) and nonadjacent to at least one neighbour of the marker if possible. Let $A$ be the set of chosen vertices.

We will show that $A$ is an asteroidal set of $H_G$ by contradiction.
Suppose not. Then there must be three vertices
$a, b, c \in A$ such that $a$, $b$ and $c$ are not pairwise nonadjacent or every $b, c$-path in $H_G$ contains a vertex of $N[a]$. 
Since $a$ is not a marker vertex, it is a vertex of $G_x$ for just one leaf $x$ of $PST_G$. Let $y$ be the neighbour of $x$ in $PST_G$ and let $m$ be the marker vertex that appears in both $G_x$ and $G_y$.
Since $a$ is nonadjacent to $m$, all of $a$'s neighbours in $H_G$ are vertices of $G_x - \{m\}$.
By similar reasoning for $b$ and $c$, we see that $a$, $b$, and $c$ are pairwise nonadjacent and no two of them have a common neighbour.
Thus, every $b,c$-path in $H_G$ must contain a vertex of $N[a]$ and, therefore, a vertex of $G_x - \{m\}$.

Let $P$ be an arbitrary $b,c$-path in $H_G$ and let $p_f$ and $p_{\ell}$ be the first and last vertices, respectively, of $P$ that are in $G_x - \{m\}$.
Because the only edges connecting $G_x$ to the rest of $H_G$ are those connecting the neighbours of $m$ in $G_x$ to the neighbours of $m$ in $G_y$, $p_f$ and $p_{\ell}$ are both neighbours of $m$ in $G_x$. Furthermore, replacing the $p_f, p_{\ell}$-subpath in $P$ with any neighbour of $m$ in $G_x$ yields a $b,c$-path in $H_G$.
Therefore, vertex $a$ must be adjacent to all neighbours of $m$ in $G_x$ as otherwise there would be a $b, c$-path not containing a node of $N[a]$. 
By the choice of $a$ this implies that all nonneighbours of $m$ in $G_x$ are adjacent to all neighbours of $m$ in $G_x$. This in turn implies that either $G_x$ is a star of which $m$ is a leaf, or $G_x$ is neither a clique nor a star and has a split. The second possibility would contradict the fact that $PST_G$ is a standard split decomposition tree; therefore $G_x$ must be a star in which the marker vertex is a leaf.

Furthermore, $m$ must be a cut vertex in $G_y$ or else there would be a $b, c$-path in $H_G$ avoiding $G_x$ and thus $N[a]$.
As mentioned in \cite{Cunningham},
any graph with four or more vertices that has a cut vertex also has a split and therefore $G_y$ is not a prime graph. Therefore, since $m$ is a cut vertex, this implies that $G_y$ is a star in which $m$ is the centre.

But then replacing $G_x$ and $G_y$ with the join composition of $G_x$ and $G_y$ in the standard split decomposition of $H_G$ yields a split decomposition of $H_G$ of which $PST_G$ is a strict refinement, contradicting the fact that $PST_G$ is a standard split decomposition tree.
Consequently, $A$ is an asteroidal set of $H_G$ and the statement follows by Lemma \ref{psipreserved}.
$\Box$

\bigskip

In the remainder of this section, we show that the asteroidal number, the polygon number, and the number of leaves in the pruned standard split decomposition tree are equal for any connected distance hereditary graph that is not a clique.
Bandelt and Mulder \cite{BM} showed that a graph is distance hereditary if and only if it can be constructed from a single vertex by repeatedly adding a leaf or substituting a vertex with a pair of twins or, equivalently, it can be transformed to a single vertex by repeatedly removing a leaf or reducing a pair of twins. 
It follows that trees and cographs are distance hereditary graphs,
and that distance hereditary graphs are circle graphs.
Hammer and Maffray \cite{HM} characterized distance hereditary graphs as graphs whose standard split decompositions contain only cliques and stars. 
A forbidden induced subgraph characterization of distance hereditary graphs was given in \cite{BM} and \cite{HM}:
a graph is a distance hereditary graph if and only if it is (house, hole, domino, gem)-free, that is, it does not contain an induced subgraph isomorphic to any of the house, holes, domino, or gem graphs, which are depicted in Figure \ref{FS}.

The next observation and technical lemma lead to a characterization of distance hereditary permutation graphs and provide a device for proving that the polygon number of a distance hereditary graph is at most the number of leaves of the pruned standard split decomposition tree. 
In particular, the lemma identifies conditions under which the fact that the pruned standard split decomposition tree of a graph is a path ensures that the corresponding graph is a connected permutation graph.

\begin{obs}\label{star-leaves}
For a connected distance hereditary graph $G$, each leaf of $PST_G$ is a star with a degree-one marker vertex. 
\end{obs}

\begin{lemma}\label{pathlemma}
Let $G$ be a connected distance hereditary graph and let $x_1, x_2, \ldots, x_j$ be the nodes of a path in $PST_G$ such that $d(x_1)=1$,
$d(x_i)=2$ for all $1<i<j$, and $d(x_j)\le 2$. For each $1 \le i \le j$, let $H_i$ 
be the graph such that $PST_G [ \{ x_1, \ldots, x_i \} ] = ST_{H_i}$. Then, for each $1 \le i \le j$,
$H_i$ is a connected permutation graph that contains at most one marker vertex and has a permutation representation in which the marker vertex, if it exists, is represented by an extreme chord.
\end{lemma}

\proof
Since $G_{x_1}$ is a clique or a star, the chords of a permutation representation for it can be permuted to make any chord extreme, so the result follows for $i=1$.
Let $i>1$ and assume that $H_{i-1}$ satisfies the lemma.
Let $\ell_{i-1}$ denote the one marker vertex in $H_{i-1}$.
If $x_i$ has degree two in $PST_G$, it contains two marker vertices: $\ell_{i-1}$ shared with $x_{i-1}$ and one other, denoted $\ell_i$, corresponding to the split between $H_i$ and $G$ minus the vertices of $H_i$. Otherwise, it contains only one marker vertex,
$\ell_{i-1}$, and in this case, $i=j$ and 
$PST_G [ \{ x_1, \ldots, x_i \} ]  = PST_G$ 
since the path $x_1, x_2, \ldots, x_j$ must be the whole tree. 
In either case, the result of the join of $G_{x_i}$ with $H_{i-1}$ will have at most one marker vertex.
Let $D$ be a permutation representation for $H_{i-1}$ in which $\ell_{i-1}$ is represented by an extreme chord. The remainder of the proof shows how to extend $D$ to a permutation representation for $H_i$ that has the desired properties.
Let $n_i$ be the number of vertices of $G_{x_i}$.

If $G_{x_i}$ is a clique, we 
replace $\ell_{i-1}$ in $D$ with $|x_i|-1$ pairwise crossing chords making $\ell_i$, if it exists, an extreme chord.

If $G_{x_i}$ is a star with $\ell_{i-1}$ the centre, then $\ell_i$, if it exists, must be a leaf of the star.
In $D$, we replace $\ell_{i-1}$ with $n_i-1$ noncrossing chords making $\ell_i$, if it exists, an extreme chord.

If $G_{x_i}$ is a star with $\ell_{i-1}$ a leaf, then if $\ell_i$ exists it could be the centre or a leaf of the star. 
In $D$, we replace $\ell_{i-1}$ with a single chord $p$ (representing the centre of $G_{x_i}$) and then add $n_i - 2$ noncrossing chords, all crossing just the chord $p$, making $\ell_i$ an extreme chord if $\ell_i$ is a leaf. Otherwise, if $\ell_i$ is the centre then it is represented by the  extreme chord $p$. 
$\Box$

\bigskip

Several characterizations of distance hereditary chordal graphs \cite{Howorka2} and distance hereditary comparability graphs \cite{DiS} are known. 
The next two theorems give characterizations of distance hereditary permutation graphs.

\begin{theorem}\label{paththm}
Let $G$ be a connected distance hereditary graph. The following are equivalent:
\begin{enumerate}
\item
$G$ is a permutation graph.
\item
$G$ is AT-free.
\item
$PST_G$ is a path.
\end{enumerate}
\end{theorem}

\proof
Statement (1) implies (2) since all permutation graphs are AT-free (see \cite{GraphClasses}).
Statement (2) implies (3) by the following justification. If the pruned decomposition tree is not a path then it has three leaves. This implies that $an(G) \ge 3$ by Lemma \ref{an>=leaves}, which contradicts the fact that $G$ is AT-free.
Finally, (3) implies (1) by Lemma \ref{pathlemma} and Lemma \ref{psipreserved}.
$\Box$

\begin{theorem}\label{DHPAT}
Let $G$ be a graph. The following are equivalent:
\begin{enumerate}
\item 
$G$ is a distance hereditary permutation graph.
\item
$G$ is a distance hereditary AT-free graph.
\item
$G$ does not contain an induced subgraph isomorphic to any of the graphs of Figure \ref{FS}.
\begin{figure}
\centering
\includegraphics[width=\linewidth]
{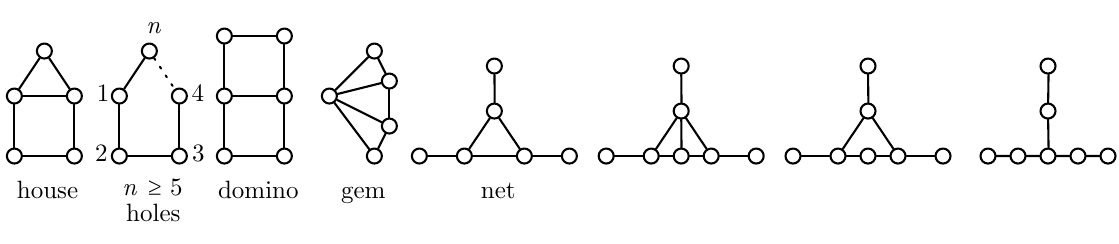}
\caption{The forbidden subgraphs for distance hereditary permutation graphs: the infinite family of holes of size greater than or equal to 5, and seven additional graphs}\label{FS}
\end{figure}
\end{enumerate}
\end{theorem}

\proof
We first show that (2) and (3) are equivalent. Since a graph is a distance hereditary graph if and only if it is (house, hole, domino, gem)-free \cite{BM} \cite{HM}, a graph is a distance hereditary AT-free graph if and only if it is (house, hole, domino, gem, AT)-free.
Gallai \cite{Gallai} gave a list of minimal ATs not containing $C_5$, consisting of four infinite families of graphs and eleven additional graphs. By inspection,
the only minimal ATs from Gallai's list that do not contain a house, hole, domino, or gem as an induced subgraph are the four graphs on the right in Figure \ref{FS}.
The equivalence of (1) and (2) follows from Theorem \ref{paththm}. 
Alternatively, the equivalence of (1) and (3) could be verified by examining the forbidden subgraphs of permutation graphs.
$\Box$

\bigskip

Theorem \ref{DHPAT} implies that the classes of distance hereditary permutation graphs, distance hereditary cocomparability graphs, and distance hereditary AT-free graphs are all the same, since permutation graphs form a subset of cocomparability graphs which in turn form a subset of AT-free graphs
(see \cite{GraphClasses}).
Di Stefano \cite{DiS} showed that a graph is a distance hereditary comparability graph if and only if it does not contain an induced subgraph isomorphic to a house, hole, domino, gem, or net (see Figure \ref{FS}). Since the net graph contains an AT, this implies that distance hereditary AT-free graphs form a subset of distance hereditary comparability graphs.

We next show how to construct a polygon representation for a connected distance hereditary graph that has the number of corners equal to the number of leaves of the pruned standard split decomposition tree of the graph.

\begin{lemma}\label{psi<=leaves}
Let $G$ be a connected distance hereditary graph and let $k \ge 2$ be the number of leaves of $PST_G$. Then $\psi(G) \le k$.
\end{lemma}

\proof
If $k=2$ then, by Theorem \ref{paththm}, $G$ is a permutation graph and $\psi(G)=2$.
Let $k>2$ and assume the lemma holds for graphs whose pruned standard split decomposition trees have fewer than $k$ leaves. 
Let $x_1, x_2, \ldots, x_j, x_{j+1}$ be the nodes of a path in $PST_G$ such that $d(x_1)=1$,
$d(x_i)=2$ for all $1<i \le j$, and $x_{j+1}$
has degree greater than two. Such a path must exist since $k>2$. By Lemma \ref{pathlemma},
$H_j$, 
the graph such that $ST_{H_j} = PST_G [ \{ x_1, \ldots, x_j \} ]$, is a connected permutation graph, contains exactly one marker vertex, and has a permutation representation in which the marker vertex is represented by an extreme chord. The marker vertex of $H_j$ is the marker vertex that $G_{x_j}$ shares with $G_{x_{j+1}}$; we refer to it as $\ell_j$.
Now let $T' = PST_G - \{ x_1, \ldots, x_j \}$.
Since $x_{j+1}$ has degree greater than two in $PST_G$, $T'$ has $k-1$ leaves and is the pruned standard split decomposition tree of a subgraph of $G$ that, by the inductive assumption, has a ($k$-1)-polygon representation.
The lemma now follows by Lemma \ref{Lcombine} and Lemma \ref{psipreserved}.
$\Box$

\bigskip

Finally, we have the following characterization of the polygon numbers of distance hereditary graphs, which generalizes Theorem \ref{paththm} to $k$-polygon graphs for $k >2$.

\begin{theorem}\label{dh}
For a connected distance hereditary graph $G$ that is not a clique, the following parameters are equal:
\begin{enumerate}
\item
$\psi(G)$
\item
$an(G)$
\item
the number of leaves in $PST_G$
\end{enumerate}
\end{theorem}

\proof
The equalities follow from Theorem \ref{k_bound_by_an}, Lemmas \ref{an>=leaves} and \ref{psi<=leaves}, and the fact that graphs that are not cliques have asteroidal number at least two.
$\Box$

\section{Upper bounds on the polygon number}\label{bounds}

In this section, we prove the following upper bounds on the polygon number of a circle graph $G$ with $n$ vertices:
$ \kappa(G)$,
$\alpha(G) +1$, and
$ \lceil n/2 \rceil$.
We note that early work on the dimension established 
the following remarkably similar upper bounds on the dimension of a comparability graph $G$ with $n$ vertices: 
$\kappa(G)$, $\alpha(G)$, and $\lfloor n/2 \rfloor$ (see \cite{Trotter}), though we leave a further examination of the relationship between these two parameters as future work.

To show that the clique cover number is an upper bound on the polygon number of a circle graph,
we rely on properties of proper circular-arc graphs.
A {\em circular-arc graph} is the intersection graph of arcs of a circle; a collection of arcs of which the intersection graph is $G$ is a {\em circular-arc representation} for $G$. 
A circular-arc graph is a {\em proper circular-arc graph} if it has a {\em proper circular-arc representation}, that is, a circular-arc representation in which no arc is contained in any other.
Proper circular-arc graphs are circle graphs since
every proper circular-arc graph has a proper circular-arc representation in which no two arcs cover the circle, and any such representation can be transformed into a circle representation in which all chords are peripheral by replacing each arc with a chord joining its endpoints (see Theorem 8.18 of \cite{Golumbic} and Observation \ref{point_in_common}). Conversely, a circle representation containing only peripheral chords can be transformed into a proper circular-arc representation by replacing each chord with one of its empty arcs. Thus,
a graph $G$ is a proper circular-arc graph if and only if it has a circle representation in which all chords are peripheral.

We now identify a subset of the chords of an arbitrary circle representation $\mathcal{C}$ such that the clique cover number of the intersection graph of the subset dictates $\rp(\mathcal{C})$.

\begin{theorem}\label{kappa-peripheral}
Let $\mathcal{C}$ be a circle representation for a graph $G$ that is not a clique.
Then $\rp(\mathcal{C}) = \kappa(G_{\mathcal{C}_{\emptyset}})$ where $\mathcal{C}_{\emptyset}$ is the set of peripheral chords in  $\mathcal{C}$ and $G_{\mathcal{C}_{\emptyset}}$ is the intersection graph of
$\mathcal{C}_{\emptyset}$.
\end{theorem}

\proof
Let  $\mathcal{C}_{\emptyset}$ be the peripheral chords of $\mathcal{C}$ where each $c_i \in \mathcal{C}_{\emptyset}$
has endpoints $e_i$ and $e_i'$.
$G_{\mathcal{C}_{\emptyset}}$ is a proper circular-arc graph since it is the intersection graph of a set of peripheral chords.
If $\mathcal{C} = \mathcal{C}_{\emptyset}$, then $G_{\mathcal{C}_{\emptyset}}$ is not a clique by the conditions of the theorem.
If $\mathcal{C} \ne \mathcal{C}_{\emptyset}$, then 
$\mathcal{C}_{\emptyset}$ contains at least two peripheral chords -- one in each arc of a non-peripheral chord of $\mathcal{C}$ --
that do not intersect each other; therefore $G_{\mathcal{C}_{\emptyset}}$ is not a clique.
Thus in either case, $\kappa(G_{\mathcal{C}_{\emptyset}}) \ge 2$.

Consider a set of $\rp(\mathcal{C})$ corners whose addition to $\mathcal{C}$ produces a $\rp(\mathcal{C})$-polygon representation for $G$.
Each empty arc of a peripheral chord contains a corner and, for each corner, the set of peripheral chords with an empty arc that  contains that corner forms a clique by Observation \ref{point_in_common}.
Therefore, $\kappa(G_{\mathcal{C}_{\emptyset}}) \le \rp(\mathcal{C})$.

In the remainder of the proof we show that 
$\rp(\mathcal{C}) \le \kappa(G_{\mathcal{C}_{\emptyset}})$.
If $G$ has a universal vertex then by Observation \ref{uv}, the chord associated with that vertex has two empty arcs and therefore $\rp(G) = 2 \le \kappa(G_{\mathcal{C}_{\emptyset}})$.
From now on, we assume that $G$ has no universal vertex.

Tucker \cite{T-SICOMP} showed that in any circular-arc representation for a circular-arc graph with clique cover number two,
there are two points on the circle such that every arc contains one or both of the points. 
Thus, if $\kappa(G_{\mathcal{C}_{\emptyset}})=2$, there are two such points with respect to the empty arcs of $\mathcal{C}_{\emptyset}$.
No empty arc of a chord $c \in \mathcal{C}_{\emptyset}$ can contain both of the points, as that would contradict that the chord with both endpoints in the nonempty arc of $c$ (which exists since $G$ has no universal vertex) contains one of the points.
Therefore, adding corners at those two points yields a $\kappa(G_{\mathcal{C}_{\emptyset}})$-polygon representation for $G_{\mathcal{C}_{\emptyset}}$, which is also a $\kappa(G_{\mathcal{C}_{\emptyset}})$-polygon representation for $G$ by Observation \ref{sat_periph}.
Therefore, $\rp(\mathcal{C}) \le \kappa(G_{\mathcal{C}_{\emptyset}})$ if $\kappa(G_{\mathcal{C}_{\emptyset}}) = 2$.

\begin{figure}
\centering
\includegraphics[scale=1]{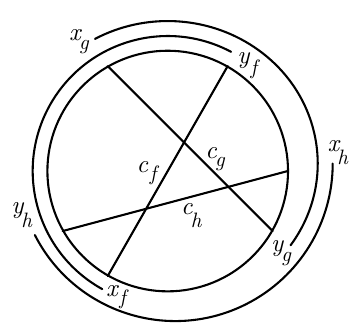}
\caption{Three pairwise crossing peripheral chords whose empty arcs (shown) do not contain a point in common}\label{ExCover3}
\end{figure}

To complete the proof, suppose that $\kappa(G_{\mathcal{C}_{\emptyset}}) \ge 3$ and consider a minimum clique partition for $G_{\mathcal{C}_{\emptyset}}$. 
If every clique in the partition corresponds to a set of chords whose empty arcs contain a point in common, then adding corners at the common points gives a $\kappa(G_{\mathcal{C}_{\emptyset}})$-polygon representation for $G_{\mathcal{C}_{\emptyset}}$ and therefore,
by Observation \ref{sat_periph},
for $G$.
Now suppose that this is not true, and so there is some clique in the partition -- corresponding to the set of chords $K$ -- such that the empty arcs of the chords of $K$ do not all contain a point in common.
Let $c_f$ and $c_g$ be chords in $K$ such that the intersection of the empty arcs of these chords does not contain an endpoint from any of the other chords in $K$.
Such a pair of chords exists since $K$ is finite.
In the remainder of this proof, we will assume without loss of generality that $(e_i, e_i')$ is the empty arc of chord $c_i$.
Then the empty arcs of $c_f$ and $c_g$ are $(e_f, e_f')$ and $(e_g, e_g')$, respectively, and we can assume that $\langle e_f, e_g, e_f', e_g' \rangle$ without loss of generality.
This means that $(e_g,e_f')$ is the arc corresponding to the intersection of the empty arcs of $c_f$ and $c_g$.
By the choice of $c_f$ and $c_g$, the empty arc of each other chord of $K$ either contains or is disjoint from the arc $(e_g,e_f')$.
Since we assumed that the empty arcs of chords in $K$ do not all have a point in common, there must be at least one chord in $K$ whose empty arc is disjoint from $(e_g,e_f')$.
Let $c_h$ be such a chord.
Since $c_h$ intersects both $c_f$ and $c_g$, it must be that 
$(e_f, e_f')$, $(e_g, e_g')$, and $(e_h, e_h')$
cover the circle and $\langle e_f, e_h', e_g, e_f', e_h, e_g' \rangle$ holds.
This situation is depicted in Figure \ref{ExCover3}.

Now consider some $c_i \in \mathcal{C}_{\emptyset} - \{ c_f, c_g, c_h \}$ 
with empty arc $(e_i , e_i' )$, the existence of which is guaranteed by the fact that 
$\kappa(G_{\mathcal{C}_{\emptyset}}) \ge 3$.
If $e_i \in (e_f, e_g)$ then $e_i' \in (e_f', e_f)$ or else the fact that $(e_f, e_f')$ is empty is contradicted.
As a result, $(e_i, e_i')$ contains $(e_g, e_f')$.
Similarly, if $e_i \in (e_g, e_h)$ then $(e_i, e_i')$ contains $(e_h, e_g')$, and if $e_i \in (e_h, e_f)$ then $(e_i, e_i')$ contains $(e_f, e_h')$.

Therefore, all chords in $\mathcal{C}_{\emptyset}$ have an empty arc containing at least one of $(e_f, e_h')$, $(e_g, e_f')$, or $(e_h, e_g')$.
Thus, we can partition $\mathcal{C}_{\emptyset}$ into three sets such that the empty arcs of all chords in the same set contain the same one of these intervals.
By Observation \ref{point_in_common}, each partition forms a clique.
Putting a corner in each interval yields a $3$-polygon representation for $\mathcal{C}$, and therefore $\rp(\mathcal{C}) \le \kappa(G_{\mathcal{C}_{\emptyset}}) $.
$\Box$

\begin{cor}\label{kappa}
For any circle graph $G$ that is not a clique,
$\gp(G) ~\le~ \kappa(G)$.
\end{cor}

\proof
$\gp(G) \le \rp(\mathcal C) = \kappa(G_{\mathcal{C}_{\emptyset}}) \le \kappa(G)$
where the first inequality is by Observation \ref{minoverallreps}, the equality is by Theorem \ref{kappa-peripheral}, and the last inequality follows from the fact that $G_{\mathcal{C}_{\emptyset}}$ is an induced subgraph of $G$.
$\Box$

\begin{cor}\label{corollary:alpha}
\label{alpha}
For any circle graph $G$,
$\gp(G) \le \alpha(G) +1$.
\end{cor}

\proof
The proof is similar to that of Corollary \ref{kappa}, and uses the fact that $\kappa(G) \le \alpha(G) + 1$ for any circular-arc graph $G$ \cite{Gavril74}.
$\Box$

\bigskip
By Corollary \ref{kappa}, cobipartite circle graphs are permutation graphs and this, combined with the fact that permutation graphs are closed under complement, implies a result of \cite{Bou2} that any bipartite graph whose complement is a circle graph is itself a circle graph.
Odd chordless cycles with five or more vertices show that the bounds of Corollaries \ref{kappa} and \ref{corollary:alpha} are tight,
while permutation graphs demonstrate that $\kappa(G)$ and $\alpha(G) +1$ can be arbitrarily larger than $\gp(G)$.

We will now show that the polygon number of a graph with $n \geq 3$ vertices is at most $\lceil n/2 \rceil$.
We note that we cannot simply add corners to an arbitrary circle representation to obtain this bound, since there are circle representations with polygon number (of the representation) greater than $\lceil n/2 \rceil$.
For example, consider the graph given by the an independent set of $n$ vertices and the circle representation for this graph in which the chords appear as an independent set in series. In this case, the polygon number of the representation is equal to $n$. Another example is the $n$-vertex graph consisting of a clique on $n/3$ vertices with two leaves attached to each vertex of the clique, for $n \ge 6$ and $n$ a multiple of 3. The polygon representation in which the leaves all form an independent set in series requires $2n/3$ corners whereas the polygon number of the graph is $n/3$.
Our proof of the $\lceil n/2 \rceil$ bound avoids this problem by starting with 
a special polygon representation, the existence of which is guaranteed by Lemma \ref{intermed}.

\begin{figure}
\centering
\includegraphics[scale=1]{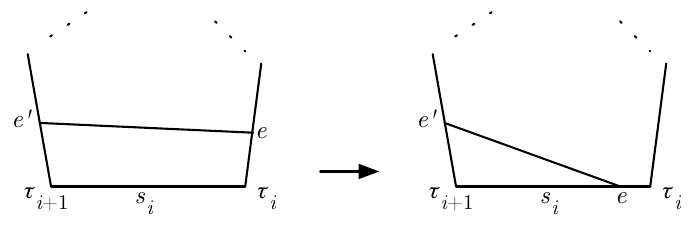}
\caption{If $e$ is close to $\tau_i$ on side $s_{i-1}$ and $e'$ is not on side $s_i$, as in the configuration on the left,
then $e$ can 
{\em slide} clockwise around $\tau_i$ resulting in the configuration on the right.}\label{Figslide}
\end{figure}

The operation of {\em sliding an endpoint around a corner} can be used to transform a $k$-polygon representation into another $k$-polygon representation for the same graph.
Let $\mathcal C$ be a $k$-polygon representation with corners $\tau_0, \tau_1, \ldots, \tau_{k-1}$ such that $\langle \tau_0, \tau_1, \ldots, \tau_{k-1} \rangle$, with sides $s_{i-1}$ and $s_i$ meeting at $\tau_i$. 
Let $c$ be a chord of $\mathcal C$ with endpoints $e$ and $e'$ such that $e$ is close to a corner $\tau_i$ on side $s_{i-1}$ and $e'$ is not on $s_i$. Then
{\em sliding $e$ clockwise around $\tau_i$} means moving $e$ from $s_{i-1}$ to $s_i$ such that $e$ is close to $\tau_i$ on $s_i$ and the relative ordering of all endpoints remains the same.
Since all chords are still satisfied in the new configuration,
the result is another $k$-polygon representation for the same graph. 
See Figure \ref{Figslide}.
Counterclockwise slides are defined analogously.

The next lemma uses the sliding operation to show that every circle graph $G$ has a special type of $\gp(G)$-polygon representation, which will be the starting point for proving the $\lceil n/2 \rceil$ bound.

\begin{lemma}\label{intermed}
\label{slide}
Let $\mathcal{C}$ be a circle representation for graph $G$ such that $\rp(\mathcal{C}) = \gp(G) = k$.
Then a set of $k$ corners $\tau_0, \ldots , \tau_{k-1}$ can be added to $\mathcal{C}$ 
--
where $\langle \tau_0, \ldots , \tau_{k-1} \rangle$,
for all $0 \le i \le k-1$, $s_{i-1}$ and $s_i$ denote the sides (in clockwise order, arithmetic modulo $k$) that meet at $\tau_i$,
and $c_i$ is the chord whose endpoint on $s_i$ is close to $\tau_i$ 
--
such that 
all chords in $\mathcal{C}$ are satisfied and
\begin{enumerate}
\item
$c_i$ is in $\tau_i$ for all $1 \le i \le k-1$, and
\item
one of the following holds:
\begin{enumerate}
\item
$c_0$ is in $\tau_0$ and
$c_0, \ldots, c_{k-1}$ form a $k$-independent set in series.
\item
$c_1, \ldots, c_{k-1}$ form a ($k$-1)-independent set in series, and if $k>2$ then all chords in $\tau_0$ intersect $c_1$ and do not intersect $c_{k-1}$, while if $k=2$ then $G$ is a clique.
\end{enumerate}
\end{enumerate}
\end{lemma}

\proof
The lemma is trivially true if $G$ is a clique since then $k=2$.
If $k=2$ and $G$ is not a clique, then let $\mathcal{C}$ be a circle representation for $G$ with $\rp(\mathcal{C})=2$, and
let $\tau_0$ and $\tau_1$ be corners which, when added to $\mathcal{C}$,
form a permutation representation for $G$.
Let $e_0$ (respectively $e_1$) be the first endpoint encountered in a clockwise traversal of side $s_0$ ($s_1$) whose chord does not correspond to a universal vertex of $G$. 
Let $c_0$ ($c_1$) be the chord whose endpoint is $e_0$ ($e_1$).
If $e_0$ and $e_1$ belong to the same chord, that chord would correspond to a universal vertex, contradicting the choices of $e_0$ and $e_1$.
Thus, $c_0 \ne c_1$.
Furthermore, $c_0$ and $c_1$ do not cross each other by the choice of $e_0$ and $e_1$ as the first endpoints satisfying the conditions.
As such, $c_0$ and $c_1$ are in corners $\tau_0$ and $\tau_1$ respectively, and they form a $2$-independent set in series.
The endpoints corresponding to chords with endpoints occurring between $\tau_0$ and $e_0$, and between $\tau_1$ and $e_1$, can then be slid as needed to ensure that $c_0$ and $c_1$ are close to each of the corners, thus satisfying conditions 1 and 2(a) of the lemma.

Now consider a $k$-polygon representation for $G$ obtained by adding $k$ corners to $\mathcal{C}$, where $k \ge 3$.
We will show how to alter the representation by sliding some of the chord endpoints around corners to obtain a $k$-polygon representation for $G$ that has the required properties.
The alteration proceeds as follows.
Visit the corners of the representation in the order $\tau_1, \tau_2, \ldots, \tau_{k-1}$; while visiting $\tau_i$ perform the following:
\begin{quote}
Let $e_i$ be the first endpoint of a chord encountered in a clockwise traversal of $s_i$ such that the other endpoint of the chord is on $s_{i-1}$.
Such a chord must exist because otherwise the corner $\tau_i$ could be removed, contradicting that the representation
contains the minimum number of corners.
Additionally, such a chord is peripheral or else we contradict the choice of $e_i$.
Now, slide all endpoints that occur between $\tau_i$ and $e_i$ (but not $e_i$) counterclockwise around the corner $\tau_i$ onto $s_{i-1}$.
\end{quote}

The above operation results in a $k$-polygon representation for $G$,
since all chord crossings remain the same as in the original representation
and all chords remain satisfied. Moreover, in the new representation $e_i$ is close to $\tau_i$ for all $1 \le i \le k-1$.

For all $i$, $1 \le i \le k-1$, let $c_i$ be the chord whose endpoint is $e_i$.
The chords $c_1, \ldots, c_{k-1}$ are distinct
since $k \ge 3$ and therefore a chord can be in at most one corner.
Suppose that $c_i$ and $c_j$ cross, for some $1 \le j < i \le k-1$.
Then $j=i-1$ and
$c_i$'s endpoint occurs before $c_j$'s endpoint in a clockwise traversal of side $s_j$.
But $c_j$'s endpoint is the closest endpoint to $\tau_j$ on side $s_j$,
a contradiction.
Therefore, the chords $c_1, \ldots, c_{k-1}$ form an independent set in series.

Now, let $e_0$ be the first endpoint of a chord in $\tau_0$ encountered in a clockwise
traversal of $s_0$. 
Such a chord must exist as otherwise $\tau_0$ could be removed.
Let $c_0$ be the chord that has $e_0$ as an endpoint and let $e'_0$ be the other endpoint of $c_0$.
Since the endpoints of $c_0$ are on sides $s_{k-1}$ and $s_0$, 
chord $c_0$ cannot be identical to any of $c_1, \ldots, c_{k-1}$.
In addition, 
$c_0$ does not cross $c_{k-1}$ since $c_{k-1}$'s endpoint on $s_{k-1}$ is close to $\tau_{k-1}$, and
$c_0$ does not cross any of $c_2, \ldots, c_{k-2}$ since it does not have an endpoint on the same side as any of those chords.
If $c_0$ does not cross $c_1$ then $c_0, \ldots, c_{k-1}$ is a $k$-independent set in series, and we can slide all endpoints that occur between $\tau_0$ and $e_0$ (but not $e_0$) counterclockwise around the corner $\tau_0$ onto $s_{k-1}$.
If $c_0$ crosses $c_1$ then, since endpoints of all chords in $\tau_0$ are clockwise of $e_0$ on $s_0$, all chords in $\tau_0$ cross $c_1$.
Finally, no chord in $\tau_0$ crosses $c_{k-1}$ since $c_{k-1}$'s endpoint on $s_{k-1}$ is close to $\tau_{k-1}$.
$\Box$

\bigskip
Circle representations for even chordless cycles satisfy 1 and 2(a) of Lemma \ref{intermed}, while representations for odd chordless cycles satisfy 1 and 2(b). 
For all $n \ge 3$, $\gp(C_n) = \lceil n/2 \rceil$.
We now show that $\lceil n/2 \rceil$ is an upper bound on the minimum number of sides in a polygon representation for any circle graph with $n \ge 3$ vertices. 

\begin{figure}
\centering
\includegraphics[scale=1]{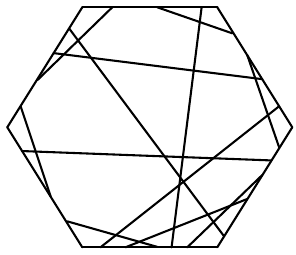}
\caption{A 6-polygon representation satisfying 1 and 2(a) of Lemma \ref{intermed}. The fact that the representation has the smallest possible number of corners follows from Theorem \ref{k_bound_by_an}. The chords that are close to the corners do not form a set of 12 distinct chords.}\label{counterexample}
\end{figure}

In the proof of the theorem, 
we begin with a $k$-polygon representation $\mathcal{C}$ for a graph $G$ such that $k = \psi(G)$ and $\mathcal{C}$ satisfies properties 1 and 2(a) of Lemma \ref{intermed}. We then show how to find 
$2k$ distinct chords in $\mathcal C$, by contradiction. 
If $2k$ distinct chords cannot be found, then after sliding some chord endpoints we can remove a corner from the representation to produce a ($k$-1)-polygon representation, which contradicts that $k = \psi(G)$.
While it is tempting to conjecture that the chords that are close to the corners form the required set, the example in Figure \ref{counterexample} shows that this collection is not necessarily a set of $2k$ distinct chords.

As part of the proof, we will construct a directed graph for which we use the following notation.
For directed graph $H=(V,E)$, we denote a directed edge from vertex $u$ to vertex $v$ as $u \rightarrow v$. The {\em indegree} (respectively, {\em outdegree}) of a vertex $v$ in $H$, denoted $indegree(v)$ (respectively $outdegree(v)$), is the number of edges $u \rightarrow v$ (respectively $v \rightarrow u$) where $u \in V$.

\begin{theorem}\label{n/2}
For any circle graph $G$ on $n \ge 3$ vertices, 
$\gp(G) \le \lceil n/2 \rceil$.
\end{theorem}
\proof
We consider only graphs without isolated vertices since removing such vertices does not affect $\gp(G)$.
The theorem is trivially true if $\gp(G)=2$, so let $G$ be a circle graph with $n \ge 3$ vertices such that $k = \gp(G) \ge 3$,
and let $\mathcal{C}$ be a $k$-polygon representation for $G$ that satisfies the conditions of Lemma \ref{slide}.
We will show that $n \ge 2k-1$.

\smallskip
\noindent
{\bf Case 1}: 
For all $0 \le i \le k-1$, the chord $c_i$ whose endpoint on $s_i$ is close to $\tau_i$, is in $\tau_i$,
and
$c_0, \ldots, c_{k-1}$ form a $k$-independent set in series (1 and 2(a) of Lemma \ref{slide}).

We will now assign two chords to each of the $k$ corners such that no chord is assigned to multiple corners. The resulting set will contain $2k$ unique chords, thus proving the theorem in this case. We begin by assigning $c_i$ to its corresponding corner $\tau_i$, meaning only one more unique unassigned chord must be assigned to each corner. 

Let us now define the set of chords $d_0, ..., d_{k-1}$ such that $d_i$ is the chord other than $c_i$ that is close to $\tau_i$. Clearly $d_i$ is in $N(c_i)$. Since $d_i \neq c_j$ for any $i \ne j$, if $d_i$ is not close to any corner other than $\tau_i$, we can assign $d_i$ to $\tau_i$ without assigning it to multiple corners. Such corners will therefore have two chords assigned to them, and so we refer to these corners as {\em easy}. We now only need to identify a unique and unassigned chord for each of the remaining {\em hard} corners.

Before assigning chords to the hard corners, we define a directed graph $H$ with vertices $x_0, \ldots, x_{k-1}$ such that for all $0 \le i,j \le k-1$ where $i \ne j$, the directed edge $x_i \rightarrow x_j$ is in $H$ if and only if $d_j \in N(c_i) - \{d_i\}$. Thus, $x_i \rightarrow x_j$ in $H$ implies that $\tau_j$ is an easy corner and $d_j$ has been assigned to $\tau_j$. Since each $d_j$ can cross at most one $c_i$ with $i \ne j$, each vertex of $H$ has indegree at most 1.
We also define, for each $0 \le i \le k-1$, the {\em range} of $\tau_i$ to be the set of chords $N(c_i) \cup \mathcal C_{\tau_i} - \{ c_i, d_i \}$ where $\mathcal C_{\tau_i}$ is the set of chords that are in $\tau_i$.

Now let $\tau_i$, where $0 \le i \le k-1$, be a hard corner. Notice that $indegree(x_i) = 0$. Therefore, since each vertex of $H$ has indegree at most 1, the vertices of $H$ that are reachable from $x_i$ induce a directed tree in $H$ that is not reachable from any other vertex corresponding to a hard corner. Moreover, for any $x_j$ that is reachable from $x_i$ where $j \neq i$, $\tau_j$ is an easy corner.

Let $P$ be a directed path in $H$ beginning at $x_i$ and ending at a vertex $x_{m}$ such that $outdegree(x_{m}) = 0$. 
To complete the proof of Case 1, we now show by contradiction that some corner corresponding to a vertex of $P$ has an unassigned chord in its range. Later we will argue that these unassigned chords together with the $d_i$'s provide enough unassigned chords to guarantee the result. Suppose that no corner corresponding to a vertex of $P$ has an unassigned chord in its range. Then, since $outdegree(x_m)=0$, $N(c_m) = d_m$. 
We will now describe a sequence of slides that results in a $k$-polygon representation of $G$ in which $c_m$ is the only chord in $\tau_m$, and $d_{m}$'s endpoints are close to $\tau_{m}$ and to another corner. We then show that this allows us to remove corner $\tau_m$, resulting in a ($k$-1)-polygon representation, which is the needed contradiction.

If $x_m = x_i$, then $d_i$'s endpoints are close to $\tau_i$ and and some other corner $\tau_{j}$ since $\tau_i$ is a hard corner. If $j = i + 1$ (\textit{ie.} $d_i $ is in $\tau_i$), slide the endpoint of $d_i$ that is close to $\tau_{i+1}$ clockwise around $\tau_{i+1}$. $d_i$ will now no longer be in $\tau_i$, regardless of whether $j = i + 1$ or not.

\begin{figure}
\centering
\includegraphics[scale=0.95]{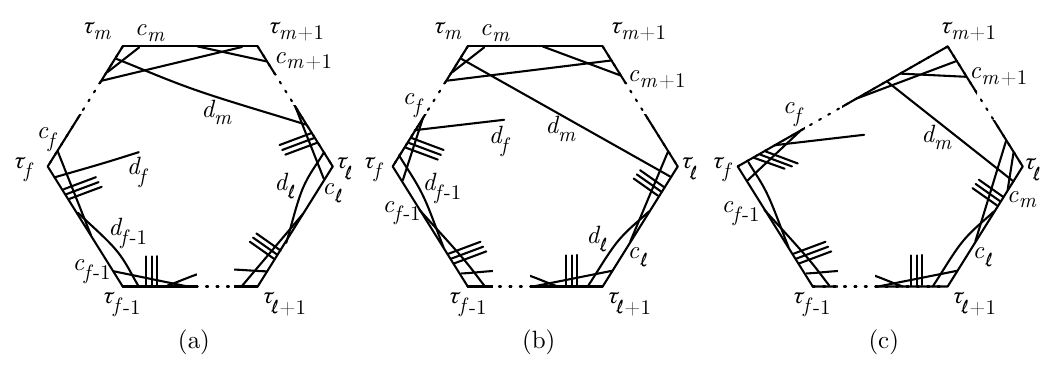}
\caption{Illustrating the proof of Theorem \ref{n/2}: (a) an initial configuration;
(b) the configuration after all of the slides have been performed; (c) the configuration after $c_m$ is moved to $\tau_{\ell}$ and $\tau_m$ is deleted.}\label{subpath}
\end{figure}

If $x_m \ne x_i$ then let $x_{\ell}$ be the predecessor of $x_{m}$ on $P$. Let $x_f$ be the first vertex of $P$ such that 
vertices of the subpath of $P$ from $x_f$ to $x_{\ell}$ correspond to corners of $\mathcal C$ that appear consecutively in a counterclockwise traversal of $\mathcal C$ starting at $\tau_f$.
That is, the subpath is $x_f \rightarrow x_{f-1} \rightarrow \cdots \rightarrow x_{\ell +1} \rightarrow x_{\ell}$ where $f$, $f-1$, \ldots , $\ell+1$, $\ell$ is a sequence of decreasing and consecutive natural numbers (arithmetic modulo $k$). 
Notice that by the selection of $f$, $d_j$ is in corner $\tau_j$ for all $\ell \leq j < f$.
See Figure \ref{subpath}(a) for an example. 

We will now slide some chord endpoints clockwise, as shown in Figure \ref{subpath}(b).
By the choice of $f$, either
$x_f = x_i$ and thus $d_f$ is close to both $\tau_f$ and some other corner, or
$d_f \in N(c_j)$ where $j \ne f+1$ and therefore $d_f$ is not in the corner $\tau_f$.
By the definition of $P$ and the choice of $f$, for all $q$ such that $\ell \le q <f$:
$d_q \in N(c_{q+1})$ and therefore $d_q$ is in $\tau_q$. 
For $j$ from $f$ down to $\ell$:
slide the chord endpoints in the empty arc of $c_j$ that are between $d_{j-1}$'s endpoint (or $d_m$'s endpoint if $j=\ell$) and $\tau_j$, including $d_{j-1}$'s endpoint (or $d_m$'s endpoint if $j=\ell$), clockwise around $\tau_j$. If any of the corresponding chords is in $\tau_j$, then first slide its other endpoint clockwise around $\tau_{j+1}$. 

After each slide of an endpoint according to this procedure, the chords will still be satisfied. 
The $d_j$ chords will be satisfied after each slide because at the moment just before $d_j$'s endpoint slides around $\tau_{j}$, 
$d_j$ is not in $\tau_j$. This is because
$d_f$ is not in $\tau_f$ by the choice of $f$, and for each other $d_j$, its other endpoint has already been slid clockwise around $\tau_{j+1}$.
By the assumption that no chord corresponding to a vertex of $P$ has an unassigned chord in its range, each other chord whose endpoint slides around $\tau_j$ has an endpoint close to another corner.
If that other corner is not $\tau_{j+1}$ then the chord remains satisfied after the slide; if the other corner is $\tau_{j+1}$ then the chord's endpoint close to $\tau_{j+1}$ is first slid around $\tau_{j+1}$ and the chord remains satisfied after each slide.
Since the chord endpoints remain in the same relative order, the resulting configuration is therefore still a $k$-polygon representation of $G$.
Clearly, each $c_j$ remains in $\tau_j$ since the endpoints of $c_j$ are not changed.
Additionally, after sliding endpoints around $\tau_j$, $d_{j-1}$ (or $d_m$ if $j=\ell$)
is close to $\tau_j$ and to $\tau_{j-1}$ (or $\tau_m$ if $j=\ell$) and is not in $\tau_{j-1}$ (or $\tau_{m}$ if $j=\ell$).

Finally, after all of these slides, $d_m$ is close to $\tau_m$ and to another corner, and $d_m $ is not in $ \tau_{m}$. Since $outdegree(x_m) = 0$, $N(c_m) = \{d_m \}$.
If there is a chord in $\tau_{m}$ other than $c_m$, then it is in the range of $\tau_m$ and so by our supposition, it must have been assigned to an easy corner. Since the chord is in $\tau_m$, the only corner it could be assigned to is $\tau_{m+1}$ and therefore its endpoint is close to $\tau_{m+1}$. Now, that endpoint can be slid clockwise around $\tau_{m+1}$ resulting in a $k$-polygon representation of $G$ in which $c_m$ is the only chord in $\tau_m$. This configuration is illustrated in Figure \ref{subpath}(b). Furthermore, $c_m$'s only neighbour is $d_m$, which has endpoints close to $\tau_m$ and to another corner. Therefore, $c_m$ can be moved to the other end of $d_m$, and $\tau_m$ can be eliminated, resulting in a ($k$-1)-polygon representation of $G$ (see Figure \ref{subpath}(c)), a contradiction.
Therefore, some corner corresponding to a vertex of $P$ has an unassigned chord in its range.

Now, for each hard corner $\tau_i$, there are at least two unassigned chords, namely $d_i$ and a chord in the range of some corner corresponding to a vertex of $H$ that is reachable from $x_i$.
Each such chord is associated with at most two hard corners. 
This is because a chord can be close to at most two corners, a chord can be in the range of at most two corners, each vertex of $H$ is reachable from at most one vertex that corresponds to a hard corner, and our use of the representation from Lemma \ref{slide} ensures that a chord can not be both in one corner and close to two other corners.
Therefore, there are at least as many unassigned chords as hard corners, and we can arbitrarily assign one unique and unassigned chord of $\mathcal C - \{ c_0, \ldots, c_{k-1} \}$ to each hard corner.
This shows that $n \ge 2k$ in this case.

\smallskip
\noindent
{\bf Case 2}: 
For all $1 \le i \le k-1$, the chord $c_i$ whose endpoint on $s_i$ is close to $\tau_i$, is in $\tau_i$. In addition,
$c_1, \ldots, c_{k-1}$ form a ($k$-1)-independent set in series, and 
all chords in $\tau_0$ 
are adjacent to $c_1$ and not adjacent to $c_{k-1}$ (1 and 2(b) of Lemma \ref{slide}).

Let $\mathcal{C}'$ be $\mathcal{C}$ with the addition of one chord in $\tau_0$ with endpoints just clockwise of the endpoint of $c_{k-1}$ on side $s_{k-1}$ and just counterclockwise of the endpoint of $c_1$ on $s_0$.
Now, $\mathcal{C}'$ satisfies conditions 1 and 2(a) of Lemma \ref{slide}.
Thus, by Case 1, it contains at least $2k$ chords and therefore $\mathcal{C}$ contains at least
$2k-1$ chords.
$\Box$

\section{Algorithmic Implications}

In \cite{HT}, Hsu and Tsai give an $O(| \mathcal{A} |)$ algorithm for computing a minimum clique cover of the intersection graph of a given circular-arc representation where $\mathcal{A}$ is the set of arcs.
A {\em minimal arc} of a set of arcs is one that does not contain any other arc in the set. 
The algorithm of \cite{HT} finds a minimum size set $S$ of minimal arcs of $\mathcal{A}$ such that every arc of $\mathcal{A}$ contains the clockwise endpoint of some arc of $S$. The collection of $|S|$ cliques of the intersection graph of $\mathcal{A}$, each of which consists of the vertices whose arcs contain the clockwise endpoint of one of the arcs of $S$, is a minimum clique cover for the intersection graph. Note that adding a non-minimal arc to $\mathcal{A}$ does not increase the clique cover number of the intersection graph, since any such arc contains a minimal arc which in turn contains the clockwise endpoint of an arc of $S$.

By Theorem \ref{kappa-peripheral}, the algorithm of \cite{HT} can be used to compute the polygon number of a circle representation in linear time, an improvement over the quadratic algorithm of \cite{ES}. 
Let $\mathcal{C}$ be a circle representation for graph $G=(V,E)$.
It is straightforward to detect whether $G$ has a universal vertex by a single scan of $\mathcal{C}$ and in that case $\rp(\mathcal{C}) = 2$ by Observation \ref{uv}.
Therefore, from  now on we consider the case in which $G$ has no universal vertex.
From $\mathcal{C}$, we construct a circular-arc representation $\mathcal{A}$ for a new graph $G'$ on $2 |V|$ vertices. The set of arcs $\mathcal{A}$ consists of both arcs of each chord of $\mathcal{C}$ where the arc endpoints are shifted slightly so that the two arcs of a single chord do not intersect but all other intersections remain the same.
By Observations \ref{uv} and \ref{point_in_common}, there is a one-to-one correspondence between the minimal arcs in $\mathcal{A}$ and the peripheral chords of $\mathcal{C}$, and $G_{\mathcal{C}_{\emptyset}}$ is an induced subgraph of $G'$. Furthermore, the vertices of $G'$ that are not in $G_{\mathcal{C}_{\emptyset}}$ are represented by non-minimal arcs in $\mathcal{A}$. Therefore by Theorem \ref{kappa-peripheral} it follows that
$\kappa(G') = \kappa(G_{\mathcal{C}_{\emptyset}}) = \rp(\mathcal{C})$, and this parameter can be computed in $O(|\mathcal{C}|)$ time by the algorithm of \cite{HT}.

The worst-case running time of the algorithm of \cite{ES} for determining whether $\gp(G) \le k$ for a given circle graph can also be improved by incorporating the algorithm of the preceding paragraph. The running time of the algorithm of \cite{ES} on a connected circle graph $G=(V,E)$ is given by the sum of the times to process each node $x$ of the standard split decomposition tree $T_G$, where a circle representation $\mathcal{C}_x$ for the graph represented by node $x$ is given. The processing time for each node is
the time to compute $\rp(\mathcal{C}_x)$ if $x$ is a prime leaf node,
$O(4^{\min \{m_x,k\}} n_x)$ if $x$ is a prime non-leaf node, and
$O(n_x)$ if $x$ is a clique or a star node, where $n_x$ is the number of vertices of $x$, and $m_x$ is the number of marker vertices of $x$. 
The overall time bound is given in \cite{ES} as $O(|V|^2 + 4^k |V|)$
since the algorithm in that paper for computing $\rp(\mathcal{C}_x)$ has running time $O(n_x^2)$. Using the algorithm of the previous paragraph improves the overall running time to 
$O(4^k |V|)$.  Although the algorithm of \cite{ES} as described above requires the input graph to be connected, disconnected graphs are easily handled within the same time bound.

The running time of the preprocessing step for the algorithm of \cite{ES} has also been improved since that paper was published. The standard split decomposition tree can be constructed and a circle representation computed for each node of the tree in 
$O((|V| + |E|) \cdot \alpha(|V| + |E|))$ time\footnote{$\alpha(|V| + |E|)$ denotes the inverse Ackermann function} by the algorithm of \cite{GPTC}, whereas previously the most efficient algorithm known for that problem was the $O(|V|^2)$ algorithm of \cite{MaSpin, Spin-Circle}.

It can be seen from the analysis of \cite{ES} as previously described, that if $m_x$ for prime nodes is bounded by a small constant then the running time of the algorithm depends on that bound rather than on $k$. In that case the algorithm, when run on the standard split decomposition tree of a circle graph $G=(V,E)$ and $k=|V|$, returns a polygon representation of $G$ with the minimum number of corners in $O(|V|)$ time. Therefore, 
$\gp(G)$ can be computed in linear time whenever the number of markers in each prime node of the standard split decomposition tree of $G$ is bounded by a constant, for example, for prime  circle graphs, the standard split decomposition trees of which have no marker vertices, and for distance hereditary graphs, the standard split decomposition trees of which have no prime nodes.

By virtue of Theorem \ref{dh}, we now have two linear time algorithms for computing $\gp(G)$ and $an(G)$ of a distance hereditary graph $G$. For connected graphs, first, the algorithm of \cite{ES} runs in linear time for distance hereditary graphs as mentioned in the previous paragraph and, second, the linear time algorithms of \cite{CMR} and \cite{Dahlhaus} for constructing the standard split decomposition tree of a graph can be combined with a simple pruning algorithm to solve the same problem. Disconnected graphs can be handled by the formula given in \cite{ES} for the polygon number of a graph in terms of the polygon numbers of its connected components, and the observation that the asteroidal number of a disconnected graph is the maximum of the asteroidal numbers of its connected components or two if each connected component is a clique.
The most efficient algorithm previously known for the asteroidal number of a distance hereditary graph is the $O(|V|^3 + |V|^{3/2} |E| )$ algorithm of \cite{aster} for the asteroidal number of $G=(V,E)$ where $G$ is in the class of HHD-free graphs, a superclass of distance hereditary graphs. 

\label{algorithms}

\section{Conclusion}

We have given bounds on the polygon number of a circle graph and shown that the polygon number of a connected distance hereditary graph that is not a clique is equal to its asteroidal number. These results lead to a forbidden subgraph characterization for distance hereditary permutation graphs and algorithms for computing polygon numbers of polygon representations and asteroidal numbers of distance hereditary graphs that are more efficient than previously known algorithms.

Finally, we mention some questions for future research. 
First, for which classes of circle graphs are the bounds of this paper satisfied with equality, and when can the polygon number be computed in polynomial time?
Clearly, if $\gp(G)$ is equal to $ \alpha(G)+1$ (or to $\lceil n/2 \rceil$) then we immediately have a polynomial time algorithm for $\gp(G)$ \cite{Gavril}.
However, having either of the other two bounds satisfied with equality does not guarantee a polynomial time algorithm for $\gp(G)$ since the clique cover problem is NP-hard for circle graphs \cite{KeilStewart} and the complexity of the asteroidal number problem on circle graphs is unknown.
Second, it seems natural to ask whether there 
are further characterizations of distance hereditary permutation graphs analogous to those of distance hereditary chordal graphs.
Third, 
it has been observed that $\gp(G) $ is at most twice the size of a minimum dominating set of $G$ (C. Paul, personal communication, 2011).
How are various other graph parameters related to the polygon number? 
As a specific example, the
dimension of comparability graphs
has some striking similarities to the polygon number.
In both cases, the graphs of parameter two are the permutation graphs.
Furthermore, the upper bounds of Section \ref{bounds} are very close to bounds on the dimension.
It would be interesting to determine the exact relationship between the two parameters.
Note that the equivalence of the classes of 2-polygon graphs, 2-dimensional comparability graphs, 2-dimensional cocomparability graphs, and permutation graphs is easily seen by considering the intersection representation of cocomparability graphs based on the dimension \cite{GRU}. 
Perhaps this observation can be extended to graphs of higher polygon number and dimension.

\bigskip
\medskip
\noindent
{\bf Acknowledgements}

\bigskip

The authors thank C. Paul for valuable discussions, and the referees for helpful comments and for pointing out the connection to comparability graph  dimension.
The authors gratefully acknowledge funding from 
NSERC (RGPIN 9217), Alberta Innovates Technology Futures, and GRAND NCE.


\bigskip
\medskip
\noindent
{\bf References}

\begin{thebibliography}{99}

\bibitem{BM}
    H.-J. Bandelt and H. M. Mulder,
    Distance-hereditary graphs,
   Journal of Combinatorial Theory Series B, 41 (1986), 182-208.
   
\bibitem{Bouchet}
   A. Bouchet,
   Reducing prime graphs and recognizing circle graphs,
   Combinatorica 7 (1987), 243-254.

\bibitem{Bou2}
   A. Bouchet,
   Bipartite graphs that are not circle graphs,
   Annales de l'institut Fourier, 49 (1999), 809-814.

\bibitem{GraphClasses}
   A. Brandst\"adt, V. B. Le and J. P. Spinrad,
   Graph Classes: A Survey, SIAM Monographs on Discrete Mathematics and Applications,
   SIAM, Philadelphia (1999).
   
\bibitem{CMR}
    P. Charbit, F. de Montgolfier, and M. Raffinot,
    Linear time split decomposition revisited,
    SIAM Journal on Discrete Mathematics, 26 (2) (2012), 499-514.

\bibitem{Cunningham}
    W. H. Cunningham,
    Decomposition of directed graphs,
    SIAM Journal on Algebraic Discrete Methods, 3 (1982), 214-228.
    
\bibitem{CunEdm}
    W. H. Cunningham and J. Edmonds,
    A combinatorial decomposition theory,
    Canadian Journal of Mathematics, 32 (1980), 734-765.
    
\bibitem{Dahlhaus}
   E. Dahlhaus,
   Parallel algorithms for hierarchical clustering and applications to split decomposition and
   parity graph recognition, 
   Journal of Algorithms, 36 (2000), 205-240.   
   
\bibitem{DiS}
  G. Di Stefano,
  Distance-hereditary comparability graphs,
   Discrete Applied Mathematics, 160 (2012), 2669-2680.

\bibitem{DCKX}
    F. F. Dragan, D. G. Corneil, E. K\"ohler and Y. Xiang,
    Collective additive tree spanners for circle graphs and polygonal graphs,
    Discrete Applied Mathematics, 160 (2012), 1717-1729.
    
\bibitem{ES}
   E. S. Elmallah and L. K. Stewart,
   Polygon graph recognition,
   Journal of Algorithms, 26 (1998), 101-140.

\bibitem{ES2}
   E. S. Elmallah and L. K. Stewart,
   Independence and domination in polygon graphs,
   Discrete Applied Mathematics, 44 (1993), 65-77.
   
\bibitem{GHS}
   C. P. Gabor, W.-L. Hsu, and K. J. Supowit,
   Recognizing circle graphs in polynomial time,
   Journal of the ACM 36 (3) (1989), 435-473.

\bibitem{Gallai}
    T. Gallai,
    Transitiv orientierbare Graphen,
    Acta Mathematica Academiae Scientiarium Hungaricae Tomus, 18 (1967), 25-66.
    
\bibitem{Gavril}
   F. Gavril, Algorithms for a maximum clique and a maximum independent set in a circle graph,
Networks, 3 (1973), 261-273.

\bibitem{Gavril74}
   F. Gavril, Algorithms on circular-arc graphs,
Networks, 4 (1974), 357-369.

\bibitem{GPTC}
   E. Gioan, C. Paul, M. Tedder and D. G. Corneil,
    Practical and efficient circle graph recognition,
     Algorithmica, 69 (2014), 759-788.

\bibitem{Golumbic}
   M. C. Golumbic,
   Algorithmic Graph Theory and Perfect Graphs, 2nd ed., Annals of Discrete Mathematics 57,
   Elsevier, 2004.
   
\bibitem{GRU}
   M. C. Golumbic, D. Rotem, and J. Urrutia,
   Comparability graphs and intersection graphs,
   Discrete Math. 43 (1) (1983) 37-46.

\bibitem{HM}
    P. L. Hammer and F. Maffray,
    Completely separable graphs,
   Discrete Applied Mathematics, 27 (1990), 85-99.
   
\bibitem{Howorka2} 
   E. Howorka,
   A characterization of ptolemaic graphs,
   Journal of Graph Theory, 5 (1981), 323--331.

\bibitem{HT}
   W.-L. Hsu and K.-H. Tsai, Linear time algorithms on circular-arc graphs, 
   Information Processing Letters, 40 (1991), 123-129.
   
\bibitem{KeilStewart}
   J. M. Keil and L. Stewart, Approximating the minimum clique cover and other hard problems in subtree filament graphs, 
   Discrete Applied Mathematics, 154 (2006), 1983-1995.

\bibitem{treewidth}
   T. Kloks,
   Treewidth, Computations and Approximations,
   Lecture Notes in Computer Science, Vol. 842, Springer, Berlin (1994).
   
\bibitem{aster}
   T. Kloks, D. Kratsch and H. M{\"u}ller,
   Asteroidal sets in graphs, 
   in Proceedings of the 23rd International Workshop on Graph-theoretic Concepts in Computer Science (WG 97), 
   Lecture Notes in Computer Science. 1335, 1997, 229-241.
   
\bibitem{KratschStewart}
   D. Kratsch and L. Stewart,
   Approximating bandwidth by mixing layouts of interval graphs,
   SIAM Journal on Discrete Mathematics, 15 (2002), 435-449.

\bibitem{MaSpin}
   T.-H. Ma and J. P. Spinrad,
   An $O(n^2)$ algorithm for undirected split decomposition,
   Journal of Algorithms, 16 (1994), 145-160.
   
\bibitem{Spin-Circle}
   J. P. Spinrad,
   Recognition of circle graphs,
   Journal of Algorithms, 16 (1994), 264-282.
   
\bibitem{Trotter}
   W. T. Trotter,
   Combinatorics and Partially Ordered Sets: Dimension Theory,
   The Johns Hopkins University Press, Baltimore (1992).
   
\bibitem{T-SICOMP}
   A. Tucker,
   An efficient test for circular-arc graphs, 
   SIAM Journal on Computing, 9 (1980), 1-24.
   
\bibitem{Unger92}
   W. Unger,
  The complexity of colouring circle graphs,
  in Proceedings of the 
  9th Annual Symposium on Theoretical Aspects of Computer Science (STACS 92), 
  Lecture Notes in Computer Science 577, 1992, 389-400.


\end{thebibliography}
\end{document}